# Tracking solid oxide cell electrode microstructural evolution during annealing by scanning 3D X-ray diffraction microscopy

A. Shukla [a#], S De Angelis[# a], J. Wright[c], Y. Zhang[b], J. Oddershede[e], H.F. Poulsen [d], J.W. Andreasen* [a]

[a] Department of Energy Conversion and Storage, Technical University of Denmark, Kgs. Lyngby 2800, Denmark
[b] Department of Civil and Mechanical Engineering, Technical University of Denmark, 2800, Kongens Lyngby, Denmark
[c] European Synchrotron Radiation Facility, 38043, Grenoble, France
[d] Department of Physics, Technical University of Denmark, Kgs. Lyngby 2800, Denmark
[e] Xnovo Technology ApS, Galoche Alle 15, 1st floor, 4600, Køge, Denmark
# Equal Contribution

**Abstract**

Ni particle coarsening is a primary degradation mechanism in Ni/YSZ solid oxide cells, limiting the lifespan of these devices. In this study, we demonstrate the use of Scanning 3D X-ray diffraction (S3DXRD) with an unprecedented spatial resolution of 100 nm, to monitor the microstructural evolution within the 3D volume of a solid oxide cell subjected to ex situ heat treatment. Unlike conventional tomography, S3DXRD combines crystallographic information with spatial maps, enabling precise identification of grain boundaries and the determination of local curvature changes in the Ni microstructure. Our study reveals that the Ni phase undergoes significant structural changes during annealing, driven by grain growth. This transformation is characterized by a reduction in local curvature, particularly in regions where grains disappear. We observe that the disappearing grains are the smallest grains in the size distribution and are often located near pores. As a result, the most notable reduction in local curvature occurs at the Ni-pore interface. The quantitative characterization of polycrystalline microstructural evolution in Ni/YSZ system provides new insights into the mechanisms of Ni particle coarsening in SOC devices, potentially guiding strategies to enhance the long-term stability of SOC devices.

**Introduction**

Solid oxide cells (SOCs) are emerging as a crucial technology in addressing climate change, offering efficient and sustainable solutions for energy conversion and storage[1]. Despite their promise, the performance and longevity of SOC devices are significantly influenced by their operating conditions, particularly the high temperatures applied and the associated mechanical stresses[2]. The state-of-the-art SOC electrode comprises of porous cermet of nickel (Ni) and yttria-stabilized zirconia (YSZ)[3]. One of the critical issues that affect the durability of Ni/YSZ electrodes is the coarsening of the Ni particles at elevated temperatures (600-1000 °C)[3]. This coarsening leads to particle agglomeration[4], a reduction in the active three-phase-boundary (TPB) density[5], and loss of Ni percolation[6]. Thus, to effectively understand the degradation mechanisms and predict the operational lifespan of Ni-YSZ electrodes, it is essential to understand the mechanisms behind the coarsening of Ni within these electrodes.
Previous efforts to understand coarsening phenomena in SOCs have yielded valuable insights, but many of these studies have been limited to 2D observations using electron microscopy[7]. Such methods cannot capture information about phase connectivity, the evolution of the 3D structure of particles, and the active TPB density. Recently, several groups have successfully studied Ni coarsening inside Ni/YSZ electrodes in 3D with a variety of advanced methods. These include scanning electron microscopy (SEM) combined with focused ion beam (FIB-SEM)[8], X-ray tomography[9] and X-ray ptychography[10]. Nelson et al.[11] used X-ray nanotomography to show that the coarsening of Ni particles is driven by a reduction in local curvature of Ni particle, corroborating similar findings reported by other research groups.[12]

However, each Ni particle has a substructure consisting of individual crystallites or grains. Iordache et al.[13] observed and quantified grain growth in monophase nanocrystalline Ni, with grain sizes ranging from 300 to 600 nm, after annealing at 600°C; conditions that closely resemble the operational environment of Ni in Ni-YSZ SOC devices. The results suggest that grain growth in Ni may significantly contribute to the microstructural evolution during SOC operation. Therefore, a comprehensive understanding of Ni particle coarsening requires examining not only the overall particle behavior but also the coarsening dynamics of the individual grains within the particles.

None of the studies mentioned above or in the literature have investigated the evolution of the polycrystalline microstructure of Ni in Ni/YSZ after annealing. The main barrier to conducting such a study in 3D has been the required spatial resolution. To capture the polycrystalline microstructural changes in Ni, with grain sizes as small as 300 nm, we need an imaging technique that offers both a suitable orientation contrast and a spatial resolution close to 100 nm or better.

While 3D Electron Backscatter Diffraction (EBSD) offers high spatial resolution and orientation contrast, it is a destructive technique, making it impossible to study the microstructural evolution in the same sample[14]. State-of-the-art X-ray diffraction imaging techniques, such as Bragg Coherent Diffraction Imaging[15] (BCDI) and dark-field X-ray microscopy[16] (DFXM), provide nanoscale resolution capable of resolving grain boundaries, however, these methods are limited by their small field of view, making them unsuitable for characterization of volumes representative of SOC devices.

Three-dimensional X-ray diffraction[17–19] (3DXRD) is a widely used technique for detailed studies of polycrystalline materials with evolving microstructures, with a spatial resolution of 1–2 μm [20–23]. Recently, a variant of the technique, scanning 3DXRD[24] (S3DXRD), was demonstrated to be capable of comprehensively characterizing polycrystalline microstructure with a spatial resolution of 100 nm[25–28].

In this study, we employ S3DXRD and X-ray fluorescence (XRF) to investigate the coarsening process of Ni at the individual grain level following annealing, focusing on a representative volume with sufficient grain statistics to enable robust statistical analysis. The primary aim is to understand how coarsening occurs within individual Ni particles and improve the understanding of curvature-driven coarsening theories at the particle level while assessing their applicability at the grain scale.

## Results

The experimental setup is schematically illustrated in supplementary information Figure S1. Simultaneous S3DXRD and XRF experiments were conducted at the nanoscope end station at the ID11 beamline of the European Synchrotron Radiation Facility (ESRF). A monochromatic beam of size 100 x 100 nm² and an energy of 42 KeV was used, with an exposure time of 5 milliseconds. The study was performed on a sub-volume of the SOC device, measuring 11*11*4 µm³, as highlighted as the yellow cylinder in Figure S1 (see Methods for more details on sample preparation). After a 3D mapping of the sample, it was annealed at 800°C for 24 hours in a reducing atmosphere (5% $H_2$, 95% Ar). The sample cooled down within 1 hour, after which the experiment was repeated on the same volume. The data were analyzed using the ImageD11 package. Additional information about the experiment and data analysis can be found in the methods section. Hereafter, the volume before annealing will be referred to as 'state 1' and the volume after annealing as 'state 2'.

**Microstructure quantification**

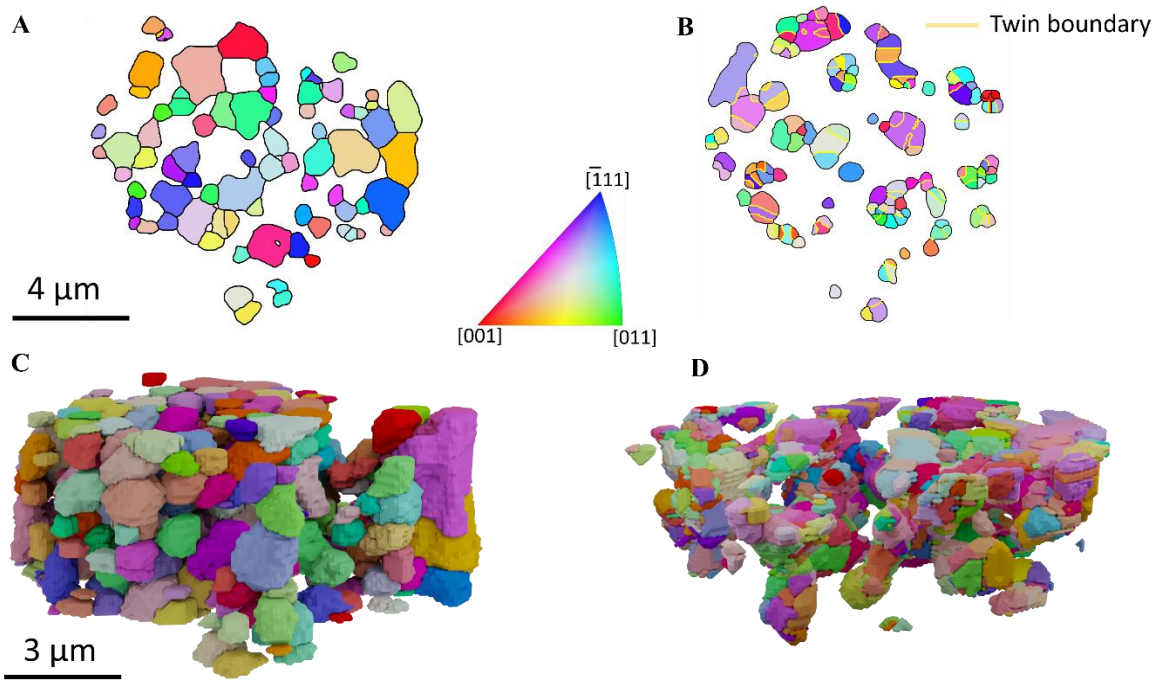

Figure 1 2D cross-section of polycrystalline microstructure for state 1 A) Only YSZ B) Only Ni C) 3D polycrystalline microstructure for state 1 C) Only YSZ. D) Only Ni. The color scale is the IPF-Z color scale as shown.

Figures 1A and 1B are visualization of 2D slices in the XY plane of the YSZ and Ni phase respectively corresponding to state 1 while Figures 1C and 1D are the visualization of corresponding 3D volumes. The microstructure is composed of three phases: Ni particles, YSZ particles and pores. The Ni-pore, YSZ-pore, and Ni-YSZ surfaces in the 3D voxelated volume were converted into a triangulated 3D mesh using Dream3D software[29] to quantify this microstructural evolution. Table 1 presents the distribution of the three-phase boundaries in the studied volume for state 1 and state 2. It is observed that the boundary areas of both the Ni-pore and Ni-YSZ interfaces significantly decrease after annealing, whereas the YSZ-pore boundary area shows minimal change. To investigate this further, the following sections focus on the evolution of the Ni polycrystalline structure, particularly at the pore and YSZ boundaries.

|  | Area of phase boundaries | | |
| --- | --- | --- | --- |
| Phase boundary type | before annealing (μm²) | after annealing (μm²) | Percentage change |
| Ni-pore | 101 | 93.68 | -7.8 |
| YSZ-pore | 202.62 | 206.82 | +2.08 |
| Ni – YSZ | 307.7 | 257.94 | -16.9 % |
| Total surface boundary area (excluding Ni-YSZ boundaries) | 302.62 | 300.5 | -0.7% |

Table 1. Distribution of Ni, YSZ and pore phase boundaries in the pre-annealed and post-annealed state of the observed volume.

**Overall evolution of the Ni microstructure**

Each Ni particle in the 3D volume is composed of multiple grains. Figures 2A and 2B show the 2D cross sections of the Ni grains in state 1 and state 2, respectively. YSZ phase has been greyed out for clarity. All grains are colored with respect to the IPF-Z color scale shown in Figure 1. The 3D visualization of Ni grain and YSZ grains are shown in the supporting information Figure S2. Twin boundaries are highlighted in yellow. After the grain segmentation procedure with a misorientation threshold of 3° (see methods), we determined that 807 grains were present in state 1. This number was reduced to 489 after annealing.

| Number | Before annealing | After annealing | Percentage change |
| --- | --- | --- | --- |
| Ni grains | 807 | 489 | -39% |
| Ni grains (twin merged) | 301 | 161 | -46 % |
| YSZ grains | 264 | 253 | 4 % |
| Ni-Ni grain boundaries | 2431 | 1223 | -49.6% |
| Ni-YSZ grain boundaries | 1611 | 1121 | -30.4% |
| Ni-Ni Σ3 boundaries | 533 | 292 | -24.5% |
| Ni-Ni Σ9 boundaries | 202 | 110 | -10% |
| Ni-Ni grain boundaries (twin merged) | 666 | 275 | -58% |
| Ni-YSZ grain boundaries (twin merged) | 1159 | 812 | -29.9% |

Table 2. Distribution of Ni and YSZ grains and grain boundaries in the pre-annealed and post-annealed state of the observed volume. Merging of twins refers to merging twinned grains in the Ni phase.

After grain tracking (see methods), 443 grain orientations from state 2 (91%) can be traced back to state 1. Also, 364 grain orientations that were found in state 1 are not found in state 2. This is direct evidence of grain growth in the Ni-YSZ cermet as the disappearing grains are being consumed by the growing grains. Regions marked R1 and R2 in Figure 2B are examples where the disappearance of grains can be observed.

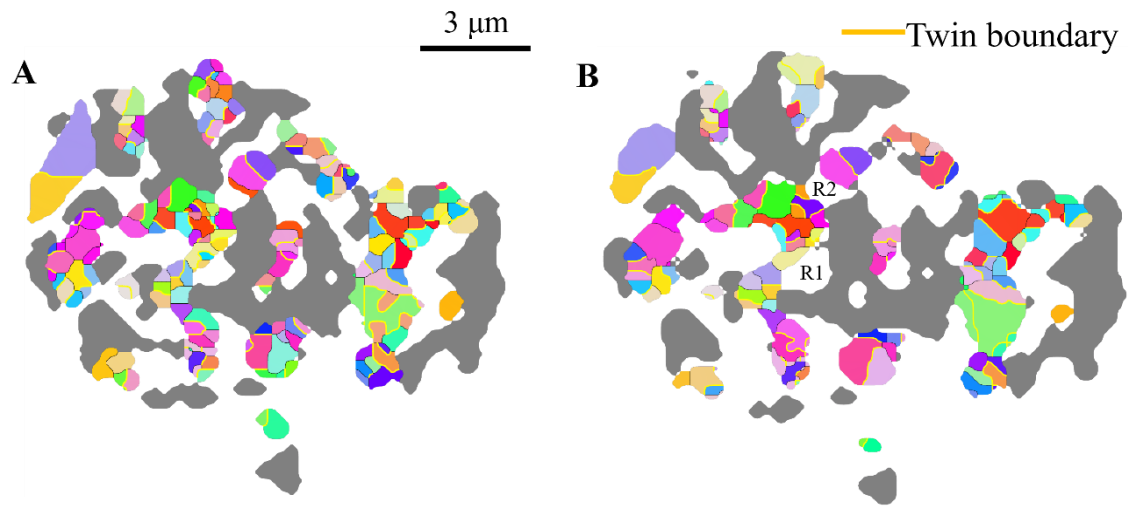

Figure 2. Visualization of 2D cross-sections of the Ni phase in the reconstructed volume in A) the pre-annealed sample (state 1) and B) the post-annealed sample (24 hours heating at 800 °C, state 2). The common color scale, IPF-Z is displayed in Figure 1. The YSZ microstructure is greyed out.

**Grain sizes, grain boundaries and texture**

During a grain growth process[30,31], some grains will disappear while other grains increase or decrease in size. In the current study, the size distribution is well represented as a log-normal distribution in both states (with a mean grain size that increases from 0.42 µm to 0.56 µm, see Figure 3A). Figure 3C shows the relation between the change in grain volume ($V_{final}-V_{initial}$) and the relative size (the ratio $D_{initial}/<D_{initial}>$) of the grain in state 1. Here $D_{initial}$ is the grain diameter in state 1, $D_{final}$ is the grain diameter in state 2, $<D_{initial}>$ is the average diameter of all grains in state 1. The higher density of grains in the first quadrant (larger-than-average size with increased diameter) suggests that big grains tend to grow during annealing, consistent with the observation in Figure 3A of an overall increase in average grain size.

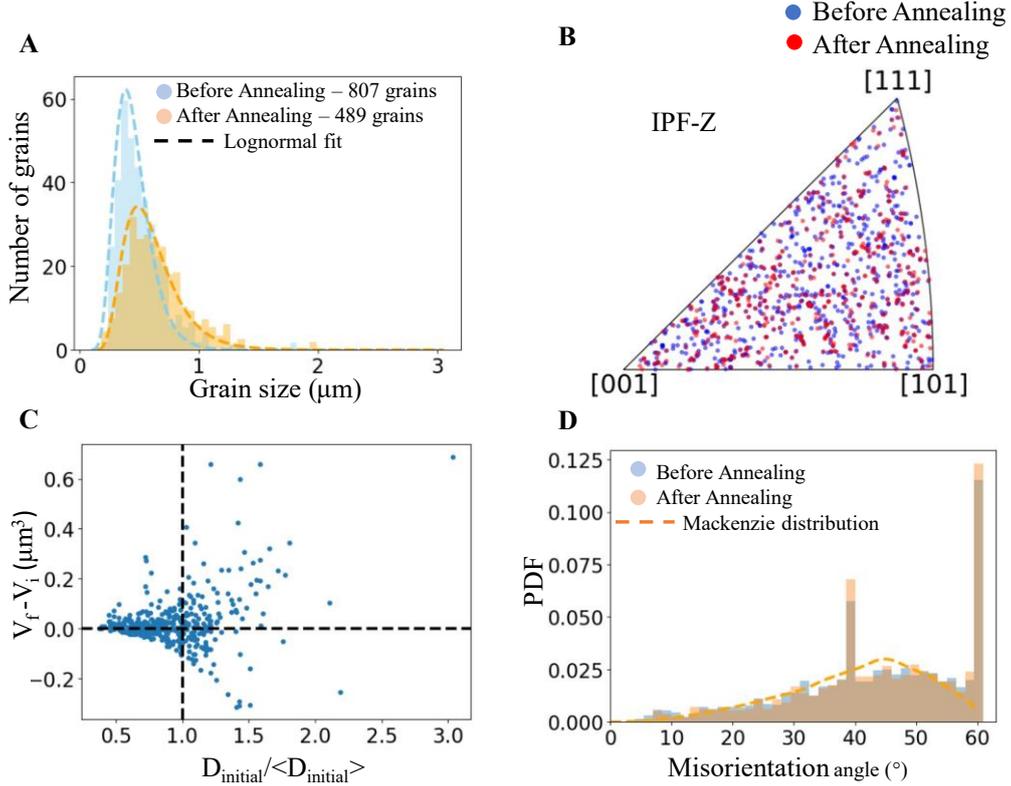

Figure 3. A) Grain size distribution of Ni before and after annealing. B) IPF-Z plot of Ni before and after annealing C) The variation in volume change as a function of the average weighted diameter of each grain, $D_{initial}/\langle D_{initial}\rangle$. D) Probability Density Function (PDF) for grain boundaries as a function of the average misorientation angle across the boundary. $\Sigma 3$ and $\Sigma 9$ twin boundaries are evident at 60° and 38.8°, respectively.

The orientation distribution in both volumes shows no significant texture. As seen in Figure 3B, the blue and red dots, representing grain orientations relative to the Z-axis, are widely dispersed within the IPF triangle. Supporting information Figure S3 shows the inverse pole figure distribution functions relative to the X-axis, with MRD values near 1, indicating minimal texture. Also, there was no noticeable orientation relationship between the YSZ phase and the Ni phase (Supporting information Figure S4A). Figure 3D shows the distribution of misorientation angles across all grain boundaries present in state 1 and state 2. The misorientation distribution is well represented by the dotted line for Mackenzie distribution as shown in the Figure 3D. It is clear that the Ni grains in Ni-YSZ are highly twinned and $\Sigma 3$ and $\Sigma 9$ twin boundaries dominate the microstructure.

It is evident from the PDF that the total fraction of $\Sigma 3$ and $\Sigma 9$ boundaries increases from state 1 to state 2, indicating a decrease in the total fraction of other boundaries. This observation aligns with the statistical data in Table 2, which shows a small reduction in twin boundaries but a much larger reduction in random-angle boundaries. This is further supported by findings from other grain growth experiments, which demonstrate that the number of low-energy twin boundaries increases relative to random-angle grain boundaries[32].

After image registration, 428 out of 489 grains in state 2 (87.5%) were paired with grains in state 1 using a 5° misorientation threshold and a 10-voxel grain centre-of-mass difference. The tracking algorithm was more effective for twin-merged grains, with 150 of 161 grains tracked successfully. This suggests that some twin grains were present only in state 2. Additionally, a few small grains rotated by more than 15°. An example of this phenomenon is shown in supporting information Figure S5. Due to the unclear cause of this rotation, these grains were classified as unmatched.

Figure 4 shows a distribution of average distance travelled by a surviving grain boundary having a minimum length of 5 voxels (See methods section for more information). It is observed that below a misorientation angle threshold of 15°, very few grain boundaries move more than 200 nm. This is in accordance with other experiments where it has been observed that low angle boundaries have lower mobilities[33]. It should be noted that that many high angle boundaries also do not move, suggesting that having a high angle grain boundary is not a sufficient condition for grain growth.

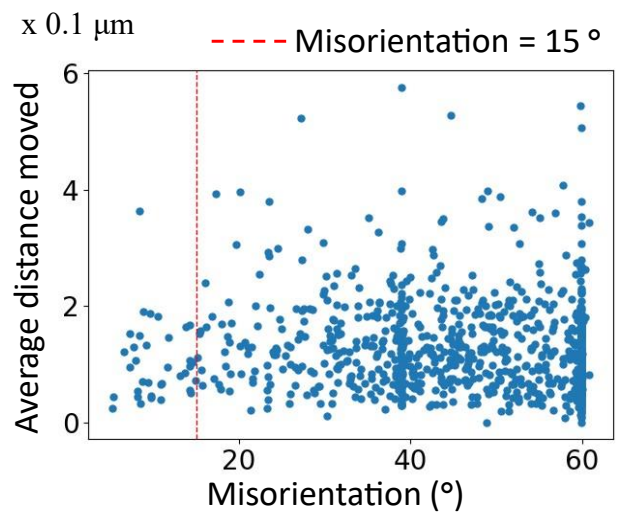

Figure 4. Histogram of the average distance travelled by 893 surviving grain boundaries as a function of grain boundary misorientation. The classical distinction between low angle and high angle boundaries is indicated by a dashed line.

**Disappearing grains**

To better understand the mechanisms behind the microstructure evolution, it is important to identify the regions in Ni/YSZ that are most impacted by grain growth, such as areas where grains disappear. Given that twin boundaries dominate the microstructure and could skew the analysis, grains sharing a twin relationship are combined into a single grain for clarity in the subsequent analysis. As shown in Figure 5A, disappearing grains tend to be smaller compared to the overall grain size distribution. This is in agreement with established grain growth theories[34].

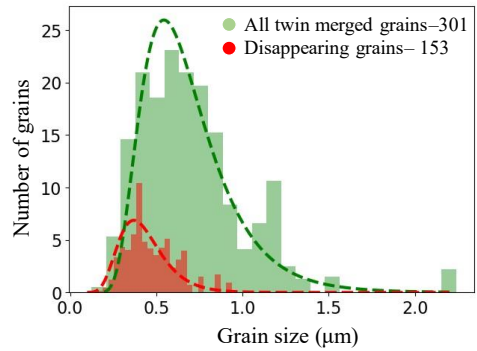

Figure 5. Grain size distributions of all twin-merged grains compared with the distribution of disappearing grains.

To evaluate the impact of the YSZ phase and pores on grain growth in Ni, we calculated the boundary fractions for each Ni grain relative to the Ni, pore, and YSZ phases, shown in Figure 6A, 6B, and 6C for Ni-Ni, Ni-YSZ, and Ni-pore boundaries, respectively. Green circles indicate surviving grains, while red circles represent grains that disappear.

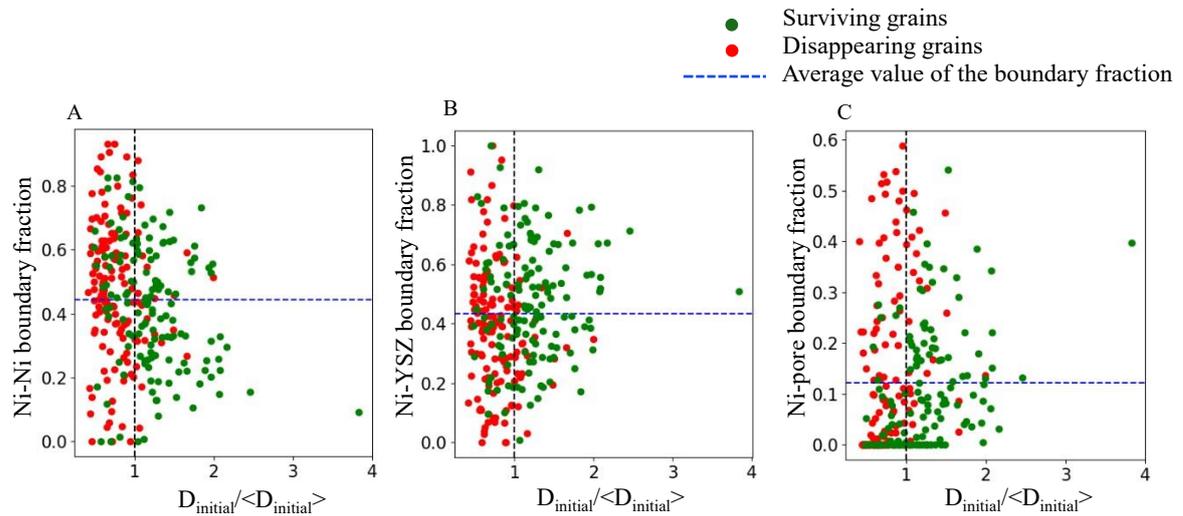

Figure 6. Scatter plots for A) Ni-Ni boundary fraction B) Ni-YSZ boundary fraction C) Ni-pore boundary fraction as a function of the average weighted diameter of each grain, $D_{initial}/<D_{initial}>$.

The black and blue dashed lines represent the average grain size and average boundary fraction for each plot type respectively. A homogeneous distribution of boundary fractions would show an equal number of grains around the blue y=constant line , representing the mean boundary fraction. However, the inhomogeneity in the distribution of surviving and disappearing grains is evident.

In all three plots, surviving grains are positioned on the right, supporting the observation from Figure 6 that smaller grains tend to disappear. Notably, Figure 6C shows very few surviving (green) grains in the second quadrant, suggesting that grains with high pore boundary fractions and low sizes are more likely to disappear (red). Similarly, Figures 6A and 6B show fewer surviving grains in the third quadrant with respect to their second quadrants, indicating that low Ni or YSZ contact correlates with lower survival rates. This is consistent with the observation in 6C that high pore fractions reduce chances of grain survival. This is in agreement with other studies where Ni-pore boundaries have proven to be a high energy boundary and hence tends to evolve[35].

**Grain growth and Ni particle coarsening**

As previously discussed, one of the major challenges to the stability of Ni/YSZ SOC devices at elevated temperatures is the reduction in the local curvature of the nickel phase. To quantify this effect, we calculated the local mean curvature at each triangle of the meshed 3D voxelated volumes using the DREAM3D software[29,36] (see Methods). As, we are only concerned about the magnitude of curvature here, the mean curvature is defined positive definite. Figures 7A and 7B present 3D visualizations of the mean curvature values for the Ni phase.

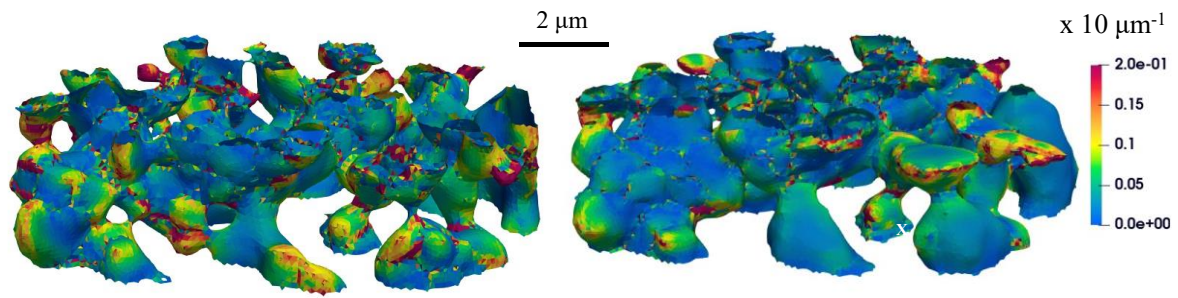

Figure 7. 3D visualization of absolute value of local mean curvature A) state 1  B) state 2.

It is important to note that the absolute curvature values here do not differentiate between a concave or a convex boundary. It is evident that in regions with significant microstructural evolution, there is a corresponding decrease in curvature values.

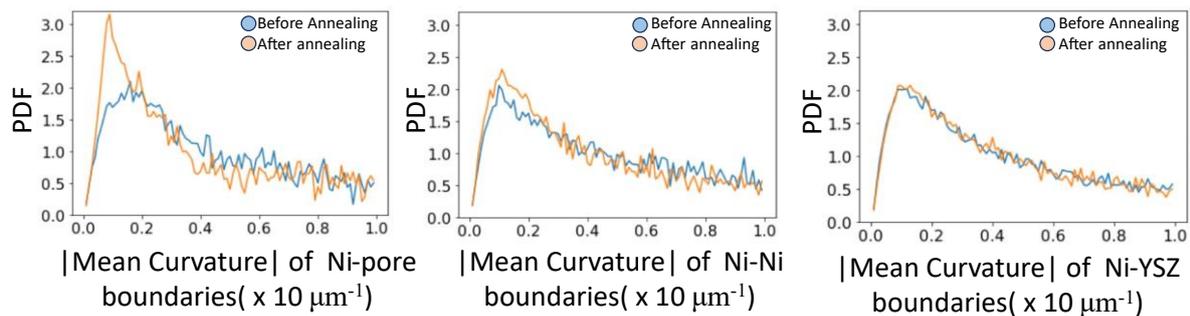

Figure 8. Grain boundary curvature distribution for all Ni grains at the two annealing time steps at the A) Ni-pore boundary B) Ni-YSZ boundary C) Ni-Ni boundary

Figure 8 shows the probability density functions (see Methods section) for curvature values for Ni grain boundaries in the two states. Whereas all boundary types show an increase in low curvature values, the curvature at Ni-pore boundaries exhibits the most significant changes during annealing. After annealing, a substantial proportion of Ni-pore boundaries display markedly low curvature values. Figure 9 is a comparison of probability density functions for disappearing grains and surviving grains calculated for all boundary types. Grains that disappeared typically had higher curvature values than the surviving grains (tail of the plots in Figure 9), while surviving grains were predominantly characterized by a larger proportion of low curvature values (peak of the two plots in Figure 9). These observations (Figure 8 and Figure 9) indicate that Ni particle coarsening is most pronounced at the Ni-pore boundaries and places where grains disappear.

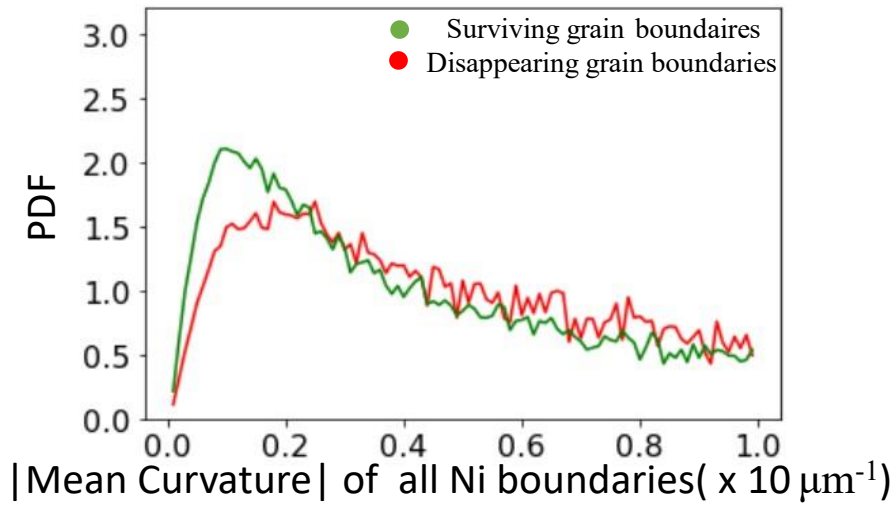

Figure 9. Grain boundary curvature comparison of surviving and disappearing grains for all types of Ni boundaries.

**Discussion**

The microstructural evolution at high temperatures in Ni/YSZ solid oxide cells remains a critical challenge for the solid oxide cell (SOC) community, as it directly affects both longevity of these devices. Our study provides comprehensive evidence that grain growth in nickel (Ni) is a key driver in the restructuring of the Ni-YSZ cermet. The primary characteristics of this grain growth are outlined as follows:

- A notable increase in grain size, accompanied by a reduction in the number of grain boundaries, was observed. This behavior is consistent with classical grain growth mechanisms, wherein the reduction of high-energy grain boundary regions leads to a lower overall energy state, pushing the system toward thermodynamic stability.

- A relative increase in the fraction of twin boundaries, known to be low-energy boundaries, was observed. This contributes to the minimization of the system's energy during annealing and suggests the preference for the twin boundaries over other grain boundaries in the evolving microstructure.

- Grain boundaries with low misorientation angle (<15°) do not show large displacements after annealing.

- Most disappearing grains are small in size, indicating a Ostwald ripening mechanism, where smaller, higher-energy grains are absorbed by larger ones. There was no correlation observed between grain boundary misorientation and the tendency to disappear. (supporting information Figure S4B.)

- Notably, these disappearing grains often exhibit higher boundary curvature and are more likely to be in contact with pores. This supports our finding that the Ni-pore boundary shows the most pronounced reduction in curvature, signifying a significant role for disappearing grains in microstructure evolution.

- Based on our results, the reduction in the Ni/pore boundary area (see Table 1) and local curvature is dominated by small grains being consumed by neighboring grains near the pores. However, the significant decline in the Ni/YSZ phase boundary area remains unclear. We attribute this decrease to thermal stresses near the Ni/YSZ interface, likely caused by differences in thermal expansion coefficients during the initial annealing in the manufacturing process. Lattice strain calculations for state 1, derived from diffraction data, revealed tensile strain on the Ni side and compressive strain on the YSZ side of the Ni/YSZ boundaries (see supporting information, Figure S6).

Our findings suggest that disappearing grains play a critical role in the microstructural evolution of Ni/YSZ during annealing. Due to their higher local curvature and proximity to pores, these grains possess larger surface areas and a higher number of triple junctions between Ni, YSZ, and pores, which are key active sites in SOC devices. Upon the disappearance of these grains post-annealing, this surface area is lost, resulting in a decrease in activity. To mitigate this effect, strategies aimed at slowing grain growth in the Ni phase are essential.

Interestingly, many of the grain growth characteristics observed in the Ni/YSZ system closely resemble those seen in pure-phase Ni metal. Insights from the metallurgical community, such as the use of alloying elements could be valuable for SOC manufacturers to explore in order to control microstructural evolution. Additionally, as low-angle grain boundaries show reduced mobility and thus less grain growth, engineering a textured Ni microstructure may be helpful. The potential for epitaxial alignment with YSZ is another avenue that could provide enhanced stability to the microstructure.

**Conclusion**

In conclusion, this study provides the first comprehensive nanoscale characterization of grain growth in Ni within a Ni/YSZ SOC electrode. Our Scanning 3D X-ray diffraction findings offer significant advancements in understanding particle coarsening behavior in SOC devices, revealing that grain growth is a major, and possibly the primary, driver of Ni phase coarsening, especially at Ni-pore boundaries. Furthermore, we offer strategies to the SOC community for improving stability, such as controlling grain growth through material engineering strategies and exploring epitaxial relationships with YSZ. These findings not only enhance our fundamental knowledge of microstructural changes in SOCs but also provide practical recommendations to extend the operational lifespan of these devices.

**Outlook**

Scanning 3D X-ray diffraction (S3DXRD) holds great promise for achieving grain boundary character distribution (GBCD) in nanocrystalline grains. However, further experimentation is necessary to establish it as the state-of-the-art method for studying nanocrystalline grain growth. These studies could provide crucial insights into the early stages of grain growth (t=0), shedding light on coarsening mechanisms that are not yet fully understood. Moreover, enhancing grain shape reconstruction methods, particularly for porous materials like Ni/YSZ SOC electrodes, is vital for improving the accuracy of grain boundary shape determination.

**Methods**

A. Sample preparation

The sample was prepared from a typical Ni-3YSZ (mol. 3%$Y_2O_3$) anode supported SOFC half-cell with a ~15 μm thick Ni-8YSZ hydrogen electrode and ~10 μm thick 8YSZ (mol. 8% $Y_2O_3$) electrolyte. The NiO-YSZ cermet was reduced for 1 h at 850 °C. Further details on the cell preparation can be found here[35]. We used the Focused Ion Beam (FIB) lift-out technique to transform the SOC electrode into a cylindrical pillar with dimensions 11*11*4 μm³. An image of the sample after FIB lift out process is shown in supporting information Figure S1. The experiment was done on a sub volume of the cylinder.

B. Experimental Setup

Scanning 3DXRD experiment was done at the ID11 beamline of the European Synchrotron Radiation Facility (ESRF) in Grenoble, France. A monochromatic beam with an energy of 42 keV was focused to a spot size of 100 nm using a pair of silicon compound refractive lenses (CRL). To cover the sample volume using the 100 nm beam, we raster scanned the sample in steps of 100 nm along y and z (refer to the coordinate system in supporting information Figure S1) with rotation as the fast axis. The area of the raster scan was 4.2 μm * 14 μm respectively. Diffraction patterns at each scan point were captured using an Eiger2 4M CdTe detector situated 126.8 mm behind the sample, with a pixel size of 75 μm and an exposure time of 0.005 s.

B. Raw data reduction to a grain map

1) Steps from analyzing raw diffraction images to reconstructing 2D grain shapes have been covered here.[27]

2) Small grains pose a challenge in reconstructing reliable sinograms because of weak scattering, making it difficult to determine grain shapes using filtered back projection accurately. This issue can be addressed in single-phase materials by giving the ownership of a voxel within candidate grains to the higher-intensity grain. However, in multiphase materials, voxel intensity is not solely dependent on a specific orientation density but is also influenced by the structure factor of adjacent phases. The presence of nearby pores further complicates the determination of precise phase boundaries.

3) To mitigate this problem, fluorescence data were simultaneously collected alongside diffraction data. The diffraction maps were initially overestimated using a gaussian filter, and a mask was applied using appropriate fluorescence intensity thresholds so that the overlap between the Ni and YSZ phases was minimal. This procedure produces a 3D mask for both the Ni and YSZ phases.

4) The segmented Euler angles from the masked voxels were processed using the open-source software DREAM.3D. A segmentation threshold of 3° was applied, and grains smaller than 5 voxels were excluded to eliminate noise from the orientation data. This methodology resulted in a grain map for one of the measured time points, with dimensions of 4 x 14 x 14 µm³ and a voxel resolution of 100 x 100 x 100 nm³.

## C. Volume Registration

Accurate alignment of the two 3D volumes is essential for effective grain tracking. To achieve this, the MATLAB function imregtform was employed to compute an affine transformation matrix. This matrix was subsequently applied using the imwarp function to align the post-annealed volume with the pre-annealed volume. The highest alignment accuracy was achieved by using fluorescence intensities as the input for the transformation.

## D. Orientation Registration

Following the derivation of the affine transformation, the rotational component can be isolated. The updated orientation matrices can then be obtained by applying the relationship

$$\mathbf{U_{new}} = \mathbf{R}\mathbf{U_{old}} \tag{1}$$

where $\mathbf{R}$ represents the rotation matrix derived from the affine transformation. $\mathbf{U_{new}}$ and $\mathbf{U_{old}}$ are the orientation matrices in the new and old frame respectively.

## E. Grain Tracking

The orientations detected in the post-annealed volume were matched to those in the pre-annealed volume within a misorientation threshold of 3°. Any grain pairs with a center of mass (COM) position difference greater than 10 voxels were excluded to avoid incorrectly matched different grains with similar orientations. Each validated grain pair was assigned a unique identifier for easy tracking and comparison.

## F. Distance travelled by a grain boundary (Figure 3)

In Figure 4, average distance travelled by a grain boundary is defined as

$$\frac{\sum d_i}{n} \tag{2}$$

Here, $i$ represents each voxel in the grain boundary, $d_i$ is the shortest distance of the voxel in the annealed state from the initial grain boundary, and $n$ is the total number of voxels forming the grain boundary. For this calculation, only grain boundaries consisting of more than 5 voxels were considered.[37]

## G. Calculation of curvature

The voxelated volume obtained after reconstruction was used to construct a surface mesh for each grain in the 3D volume. This surface mesh was then smoothed by a Laplacian smoothening algorithm using 400 iterations and a λ parameter of 0.025 to remove the staircase effect[38]. The cubic-order algorithm was used to determine the principal curvatures values κ1 and κ2 for each meshed triangle[39]. The mean curvatures were then calculated as (κ1+κ2)/2. Details about the accuracy of the curvature calculation can be found in reference[40].

H. Probability Density function (PDF)

PDF in Figures 3,8 and 9 have been calculated such that the area under the curve is 1. The distribution in Figures 8 and 9 are weighted by the curvature values to observe curvature differences at high values.

# Tracking solid oxide cell electrode microstructural evolution during annealing by scanning 3D X-ray diffraction microscopy

A. Shukla [a#], S De Angelis[# a], J. Wright[c], Y. Zhang[b], J. Oddershede[e], H.F. Poulsen [d], J.W. Andreasen* [a]

[a] Department of Energy Conversion and Storage, Technical University of Denmark, Kgs. Lyngby 2800, Denmark
[b]Department of Civil and Mechanical Engineering, Technical University of Denmark, 2800, Kongens Lyngby, Denmark
[c] European Synchrotron Radiation Facility, 38043, Grenoble, France
[d]Department of Physics, Technical University of Denmark, Kgs. Lyngby 2800, Denmark
[e]Xnovo Technology ApS, Galoche Alle 15, 1st floor, 4600, Køge, Denmark
# Equal Contribution

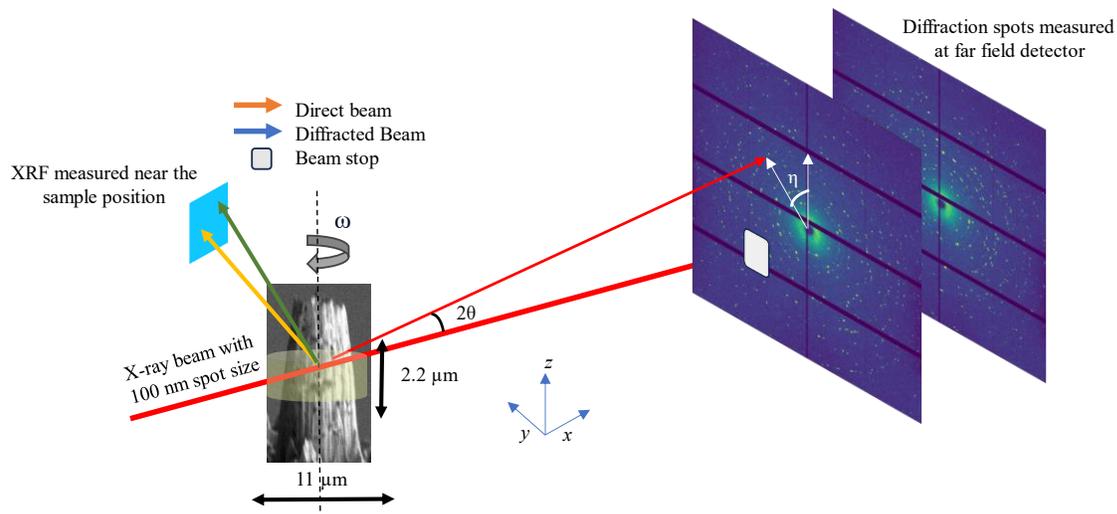

Figure S1. A) Schematic for the experiment. The X-ray beam energy is 42 KeV.

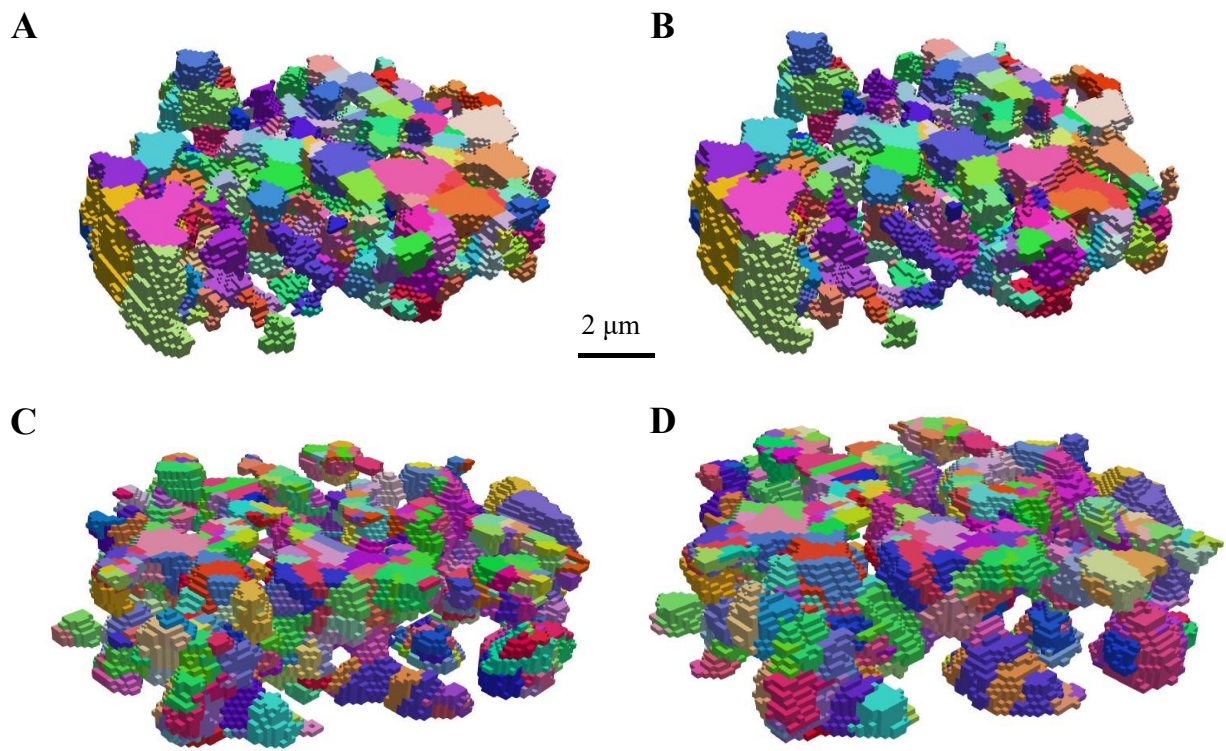

Figure S2. A) B) 3D visualization of YSZ phase for state 1 and state 2. C)D) 3D visualization of Ni phase for state 1 and state 2. It can be observed that YSZ shows minimal change in polycrystalline microstructure whereas changes in Ni phase can be observed.

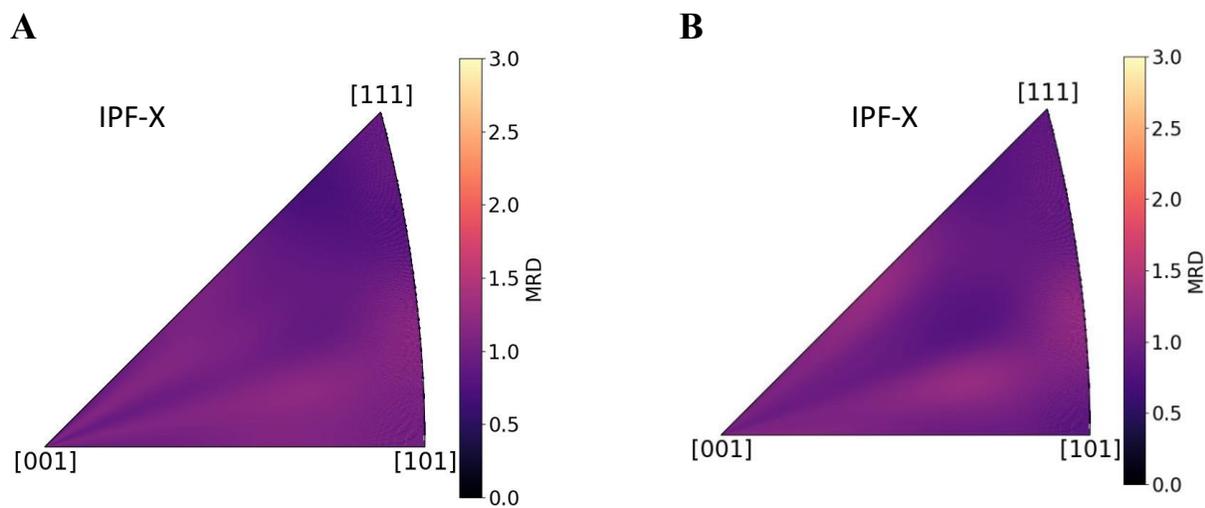

Figure S3. Inverse pole density function for A) Ni grains before annealing B) Ni grains after annealing. The value of 1 MRD (Multiple of random distribution) represents a texture less sample.

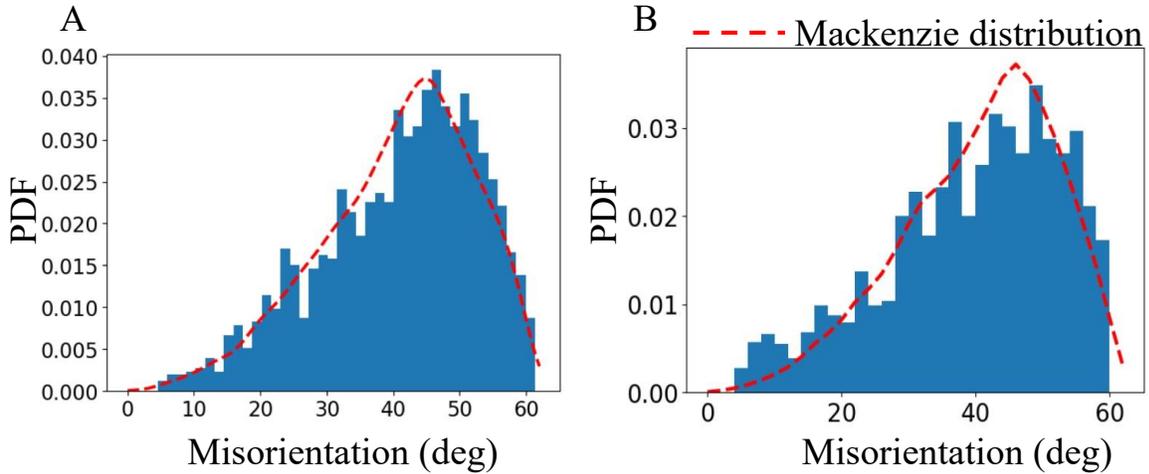

Figure S4. Distribution of grain boundary misorientation of A) Ni-YSZ grain boundaries B) grains that disappear between the state 1 and state 2. The distribution fits well with a random distribution represented by the Mackenzie fit, indicating a) There is no orientation (epitaxial relationship between Ni and YSZ, b) Disappearance of a grain after annealing does not depend on its grain boundary misorientation.

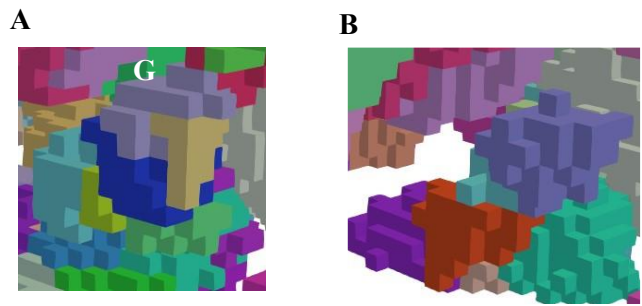

Figure S5. Grain marked G in state 1 (A) visibly changes its IPF-Z color which represents its orientation in state 2 when it consumes its neighbouring grains (B).

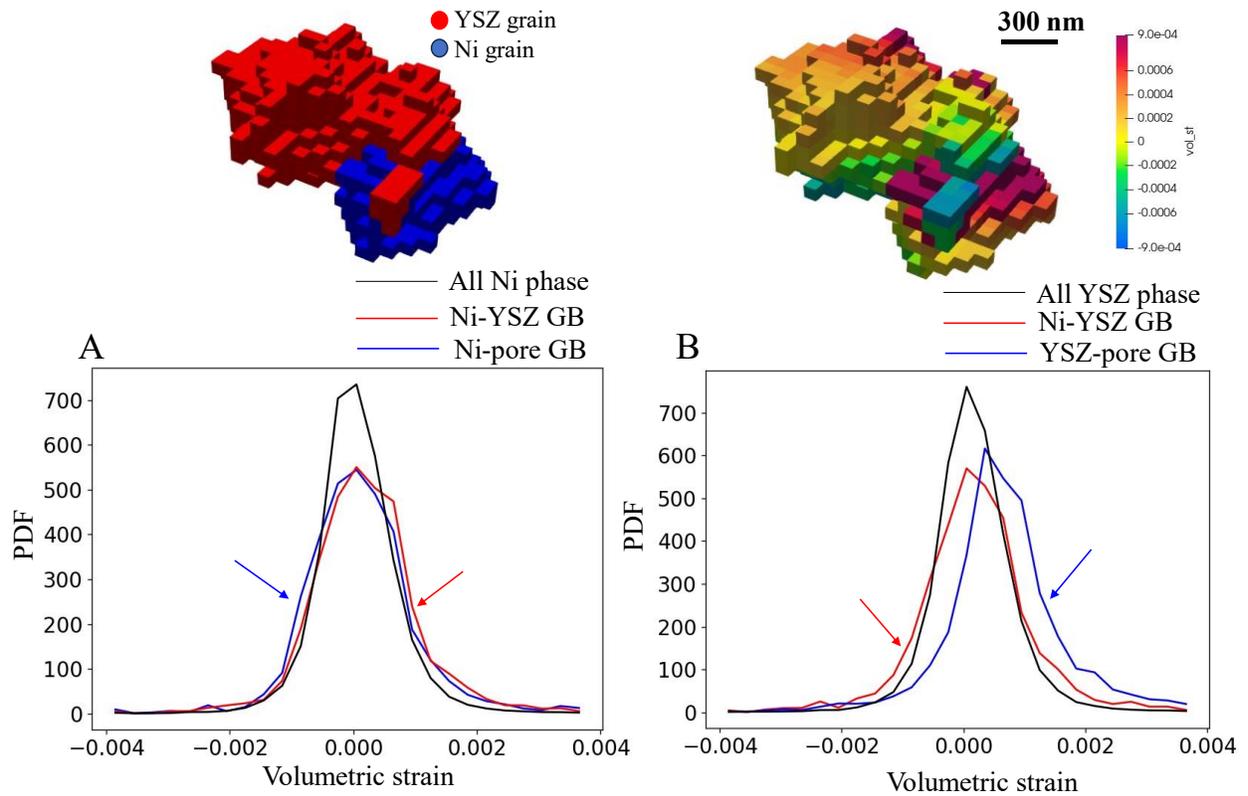

Figure S6. A. 3D map showing a YSZ grain (red) with a neighbour Ni grain (blue). B) Volumetric strain map of A. It can be observed that Ni has tensile strain while YSZ has compressive strain. C) Volumetric Strain PDF for Ni in Ni-YSZ boundaries compared with the PDF of all Ni phase  C) Volumetric Strain PDF for YSZ in Ni-YSZ boundaries compared with the PDF of all YSZ phase. It can be observed that the Ni phase in Ni-YSZ grain boundary tends to have tensile strain while the YSZ phase in Ni-YSZ grain boundary tends to have compressive strain.